\documentclass{v23tmex}

\usepackage{epsfig}
\usepackage[usenames,dvipsnames]{color}
\usepackage{slashed}

\usepackage{amssymb,cite,graphicx}
\usepackage{amsmath,bm,bbm}
\usepackage{amsfonts}

\def\be{\begin{equation}}
\def\ee{\end{equation}}
\def\bea{\begin{eqnarray}}
\def\eea{\end{eqnarray}}

\newcommand{\Photo}{\includegraphics[height=35mm]{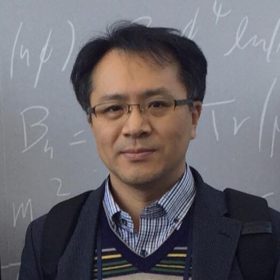}}

\begin{document}
\vspace*{4cm}
\title{Self-resonant Dark Matter}

\author{Hyun Min Lee}

\address{Department of Physics, Chung-Ang University, 84 Heukseok-ro,\\
Seoul 06974, Korea}

\maketitle\abstracts{
We present a review on the self-resonant dark matter scenarios where multiple components of dark matter give rise to a resonant condition in the $u$-channel diagrams for their comparable masses. In this case, there is no need of lighter mediators for enhancing the self-scattering and annihilation cross sections for dark matter. We discuss  the velocity-dependent self-scattering for the small-scale problems, the relic density of self-resonant dark matter, and the observable signatures in indirect and detection experiments.
}

\section{Introduction}

The observed galaxy rotation curves below $1\,{\rm kpc}$ show a deviation from the prediction in the N-body simulation with cold dark matter. The former implies the cored profile of the dark matter density at galaxies while the latter leads to the cuspy profile. This is called the core-cusp problem \cite{SIDM}. Other related problems at small scales include too big to fail problem and diversity problem, etc \cite{SIDM}.

Self-interactions of dark matter solve the core-cusp problem, because the momentum transfer from dark matter particles towards the center of galaxies makes the dark mater profile cored. However, there is no convincing evidence for self-interactions from galaxy clusters such as Bullet cluster where dark matter moves faster than in galaxies. Thus, we need to make self-interactions velocity-dependent in order to be consistent with galaxy clusters \cite{SIDM}. 

When there is a long-range force between dark matter particles with mass $m_{\rm DM}$ due to a light mediator with mass $m_{\rm med}$, it is possible to make self-interactions velocity-dependent and enhanced due to the non-perturbative effects as far as the velocity of dark matter $v$ satisfies $m_{\rm DM} v\gtrsim m_{\rm med}$. However,  the dark matter annihilation into a pair of light mediators is necessarily enhanced by the Sommerfeld effects, so it is strongly constrained by indirect detection experiments, such as Cosmic Microwave Background (CMB) anisotropies, Fermi-LAT, AMS-02, etc, if the mediator particle decays into light charged fermions in the Standard Model (SM). 

In this article, we present a review on the new possibility that an effective long-range force emerges as a consequence of the $u$-channel resonance \cite{hmlee1,hmlee2} in the case where dark matter is composed of two or more than two components with comparable masses.

\section{Bethe-Salpeter equation for $u$-channel scattering}

\begin{figure}
\begin{center}
\includegraphics[width=0.8\textwidth,clip]{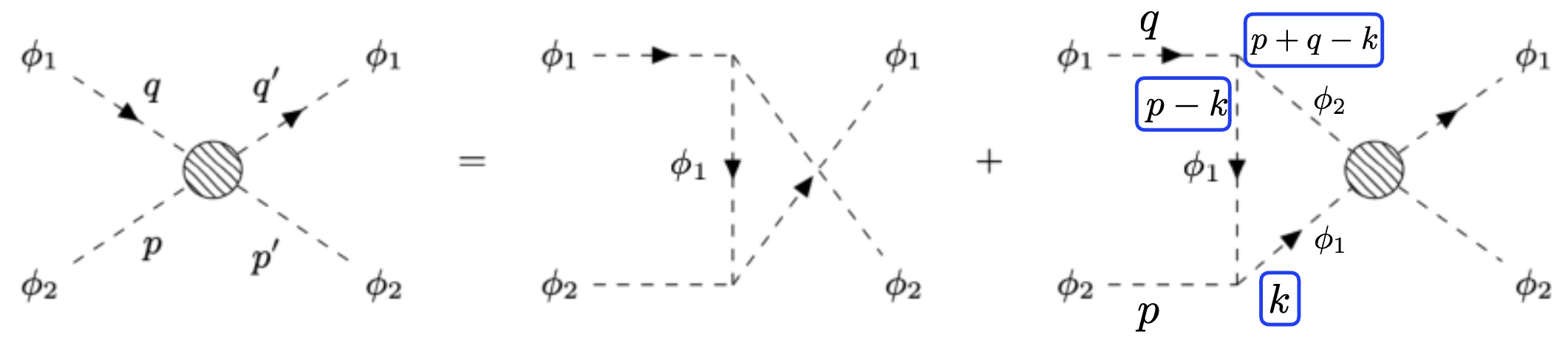} 
\end{center}
\caption[]{Non-perturbative Feynman diagrams for the $u$-channel scattering process \cite{hmlee1}.}
\label{fig:4pt}
\end{figure}

We consider a simple model for two-component dark matter with  a complex scalar $\phi_1$ and a real scalar $\phi_2$. The interaction Lagrangian for dark matter is taken to
\bea
{\cal L}_{\rm int} = -2g m_1 \phi_2 |\phi_1|^2.  \label{SRDM}
\eea
Then, as shown in the first diagram in the right of Fig.~\ref{fig:4pt}, the co-scattering, $\phi_2(p)\phi_1(q)\to\phi_2(p') \phi_1(q')$, takes place, due to the  $u$-channel exchange of the dark matter particle $\phi_1$, and the corresponding tree-level scattering amplitude is given by
\bea
{\tilde\Gamma}_u(p,q; p',q')= \frac{4g^2 m^2_1}{|{\vec p}-{\vec q}'|^2+m^2_1-\omega^2}
\eea
where $\omega=p_0-q'_0$ is the energy transfer between $\phi_1$ and $\phi_2$. Even in the non-relativistic limit for dark matter, the energy transfer does not vanish for $m_2\neq m_1$ because $\omega\simeq m_2-m_1\neq 0$, so the interaction between dark matter particles is not instantaneous. In particular, if we take $m_2=2m_1$, the $u$-channel amplitude becomes
\bea
{\tilde\Gamma}_u(p,q; p',q')= \frac{4g^2 m^2_1}{|{\vec p}-{\vec q}'|^2},
\eea
mimicking the exchange of an effectively massless mediator due to the lighter dark matter particle $\phi_1$.

In the limit of $m_2=2m_1$, it is necessary to make the resummation of $u$-channel ladder diagrams. From the non-perturbative Feynman diagrams for the $u$-channel scattering in Fig.~\ref{fig:4pt}, the recursive relation for the non-perturbative $u$-channel scattering amplitude, $\Gamma(p,q;p'q')\equiv \Gamma(p,q)$, is given \cite{hmlee1} by
\bea
i\Gamma(p,q) = i{\tilde\Gamma}(p,q) -\int \frac{d^4 k}{(2\pi)^4} {\tilde\Gamma}(p,q;p+q-k,k) G_1(k) G_2(p+q-k) \Gamma(p+q-k,k) \label{nonpert}
\eea
where $G_1(p), G_2(p)$ are the tree-level Feynman propagators for $\phi_1$ and $\phi_2$, respectively, given by
\bea
G_1(p)=\frac{i}{p^2-m^2_1+i\epsilon}, \qquad G_2(p)=\frac{i}{p^2-m^2_2+i\epsilon}.
\eea

Ignoring the tree-level amplitude in the first term of eq.~(\ref{nonpert}) and multiplying $G_2(p) G_1(q)$ to both sides of eq.~(\ref{nonpert}), we get
\bea
i\chi(p,q) \simeq -  G_2(p) G_1(q)\int \frac{d^4 k}{(2\pi)^4} {\tilde\Gamma}(p,q;p+q-k,k) \, \chi(p+q-k,k). \label{BS}
\eea
Here, we defined the Bethe-Salpeter(BS) wave function for $\phi_1$ and $\phi_2$ as
\bea
\chi(p,q) \equiv G_2(p)G_1(q) \Gamma(p,q)
\eea
and the above equation is the generalized BS equation for two particles with unequal masses. 

We redefine the momenta in terms of the center of momentum $P$ and the relative momentum $Q$ by
\bea
P=\frac{1}{2} (p+q), \quad Q=\mu\bigg(\frac{p}{m_2}-\frac{q}{m_1} \bigg),
\eea
with $\mu=m_1 m_2/(m_1+m_2)$ being the reduced mass for the $\phi_1-\phi_2$ system.
Then, noting that $\chi(p,q)={\tilde\chi}(P,Q)$ and using the BS wave function in momentum space by
\bea
{\widetilde\psi}_{BS}({\vec Q}) =\int \frac{dQ_0}{2\pi}\, {\widetilde\chi}(P,Q), 
\eea
the BS equation in eq.~(\ref{BS}) becomes
\bea
i{\widetilde\psi}_{\rm BS}({\vec Q}) = -\int \frac{dQ_0}{2\pi} \, G_2\bigg(Q+\frac{2\mu}{m_1}\,P\bigg) G_1\bigg(-Q+\frac{2\mu}{m_2}\,P\bigg) \int\frac{d^3 l}{(2\pi)^3}\, U(|{\vec Q}'|)\,{\widetilde\psi}_{BS}({\vec l}) \label{BS0}
\eea
where $U(|{\vec Q}'|)$ with ${\vec Q}'\equiv \sqrt{\frac{m_1}{m_2}}{\vec Q}+\sqrt{\frac{m_2}{m_1}} {\vec l}$ is the tree-level $u$-channel amplitude obtained after the shift of the loop momentum with $k=-l+\frac{2\mu}{m_2}\,P$, as follows,
\bea
{\tilde\Gamma}(p,q;p+q-k,k) =\frac{4g^2 m^2_1}{|{\vec Q}'|^2+ m_2(2m_1-m_2)}\equiv U(|{\vec Q}'|).
\eea

Taking the center of mass coordinates for which $P=\frac{1}{2}(m_1+m_2+E,0)$ with $E$ being the total kinetic energy, we find the non-relativistic limits of the Feynman propagators in eq.~(\ref{BS0}) as
\bea
G_2\bigg(Q+\frac{2\mu}{m_1}\,P\bigg) &\simeq& \frac{i}{2m_2\big(Q_0+\frac{\mu}{m_1}\,E -\frac{{\vec Q}^2}{2m_2}\big)+i\epsilon},\\
G_1\bigg(-Q+\frac{2\mu}{m_2}\,P\bigg) &\simeq& \frac{i}{2m_1\big(-Q_0+\frac{\mu}{m_2}\,E -\frac{{\vec Q}^2}{2m_1}\big)+i\epsilon},
\eea
and perform the contour integral for $Q_0$,
\bea
&&-\int \frac{dQ_0}{2\pi} \,G_2\bigg(Q+\frac{2\mu}{m_1}\,P\bigg) G_1\bigg(-Q+\frac{2\mu}{m_2}\,P\bigg) \nonumber \\
&\simeq& \frac{1}{2m_1m_2}\int \frac{dQ_0}{2\pi}\,  \frac{1}{\big(Q_0+\frac{\mu}{m_1}\,E -\frac{{\vec Q}^2}{2m_2}+i\epsilon\big)}\,\frac{1}{\big(Q_0-\frac{\mu}{m_2}\,E +\frac{{\vec Q}^2}{2m_1}+i\epsilon\big)} \nonumber \\
&=& \frac{i}{4m_1m_2}\, \frac{1}{E-\frac{{\vec Q}^2}{2\mu}}.
\eea
As a result, we can rewrite the BS equation in eq.~(\ref{BS0}) in the following form, 
\bea
\bigg(\frac{{\vec Q}^2}{2\mu}-E \bigg) {\widetilde\psi}_{\rm BS}({\vec Q}) =\frac{1}{4m_1m_2} \int\frac{d^3 l}{(2\pi)^3}\,  U(|{\vec Q}'|) \,{\widetilde\psi}_{BS}({\vec l}). \label{BS1}
\eea
Making a Fourier transformation of the BS wave function to position space,
\bea
\psi_{\rm BS} ({\vec x})=\int \frac{d^3{\vec Q}}{(2\pi)^3}\, e^{i{\vec Q}\cdot {\vec x}} \, {\widetilde\psi}_{\rm BS}({\vec Q}),
\eea
we finally recast eq.~(\ref{BS1}) into a Schr\"odinger-like equation \cite{hmlee1},
\bea
-\frac{1}{2\mu} \nabla^2\psi_{BS}({\vec x}) + V(\vec x) \psi_{BS}(-\beta {\vec x}) =E \psi_{BS}(\vec x) \label{Scheq}
\eea
where
\bea
V(\vec x)=-\frac{\alpha}{r}\, e^{-Mr}.
\eea
Here, the fine-structure constant for dark matter self-interaction is given by $\alpha\equiv g^2/(4\pi)$, and the effective mass for the Yukawa-type potential is given by
\bea
M\equiv m_2\sqrt{2-\beta}, \qquad \beta \equiv \frac{m_2}{m_1}.
\eea

We remark several important consequences.  First, the dark matter exchange in the $u$-channel process leads to an Yukawa-type potential for two-component dark matter with an effectively massless mediator for $m_2=2m_1$, so we can realize a long-range force without introducing light mediators.  In general, we can take $m_2< 2m_1$ for a real effective mass, forbidding the decay process, $\phi_2\to \phi_1\phi^*_1$, kinematically, and allowing for multi-component dark matter with $\phi_1$ and $\phi_2$.  Secondly, the potential term in the Schr\"odinger-like equation in eq.~(\ref{Scheq}) has the argument of the BS wave function flipped in sign by ${\vec x}\to -\beta{\vec x}$, so the effective potential is attractive for partial waves with even $l$ and repulsive for partial waves with odd $l$ where $l$ is the angular momentum quantum number \cite{hmlee1}. When the $u$-channel scattering is dominated by the $s$-wave with $l=0$, the effective potential is attractive, playing a role of enhancing the scattering cross section as in the case with a light mediator in the $t$-channel self-scattering.

\section{Self-scattering cross section for SRDM}

Writing the BS wave function in spherical coordinates as
\bea
\psi_{\rm BS}(-\beta {\vec x})= (-1)^l R_l(\beta r) Y^m_l(\theta,\phi),
\eea
the radial equation for $R_l(x)=u_l(x)/x$ with $x=\frac{1}{2} \mu\alpha r$ is given by
\bea
\bigg(\frac{d^2}{dx^2}-\frac{l(l+1)}{x^2}\bigg)u_l(x)  +\frac{4 e^{-cx}}{\beta x} (-1)^l u_l(\beta x) +a^2 u_l(x)=0
\eea
with $a=2v /\alpha$ and $c=2M/(\mu\alpha)$. 
Making a change of variables by $x=e^{-\rho}$, for $\beta=2$, we can rewrite the radial equation for ${\tilde u}(\rho)\equiv u(e^{-\rho})$ in a form of delay differential equation \cite{hmlee1,hmlee2},
\bea
{\tilde u}^{\prime\prime}_l(\rho) +{\tilde u}'(\rho) +2(-1)^l  e^{-\rho} {\tilde u}_l(\rho-\ln 2) +a^2 e^{-2\rho} {\tilde u}_l(\rho)=0 \label{radial}
\eea
where ${\tilde u}=\frac{du}{d\rho}$, etc.

We remark that the potential term in eq.~(\ref{radial}) has the argument shifted by $-\ln 2$, so we would need the information on the wave function in the "past" in order to set up the boundary condition at $x=0$ or $\rho=+\infty$. Instead, we take the boundary condition at $x=\infty$ or $\rho=-\infty$ where the plane-wave limit exists and read out the wave function at $x=0$ or $\rho=-\infty$ by the evolution of the wave function. Namely, the boundary conditions for the $s$-wave are imposed in the following \cite{hmlee1},
\bea
{\tilde u}_0(\rho)&\longrightarrow& \frac{1}{a} \,\sin(a^{-\rho}+\delta_0), \quad \rho\to -\infty, \label{bc1} \\
{\tilde u}_0(\rho)&\longrightarrow& A\,e^{-\rho}, \qquad\quad \rho\to +\infty \label{bc2}
\eea
where $\delta_0$ is the phase shift and $A$ is constant. Then, the total cross section with $s$-wave dominance is given by
\bea
\sigma=\frac{4\pi}{k^2}\,\sin^2\delta_0, \label{selfcross}
\eea
with $k=\mu v$ being the relative momentum.
Moreover, the Sommerfeld factor is given by $S_0=A^2$. The boundary conditions  in eqs.~(\ref{bc1}) and (\ref{bc2}) are generalized for higher partial waves \cite{hmlee2}.

\begin{figure}
\begin{center}
\includegraphics[width=0.4\textwidth,clip]{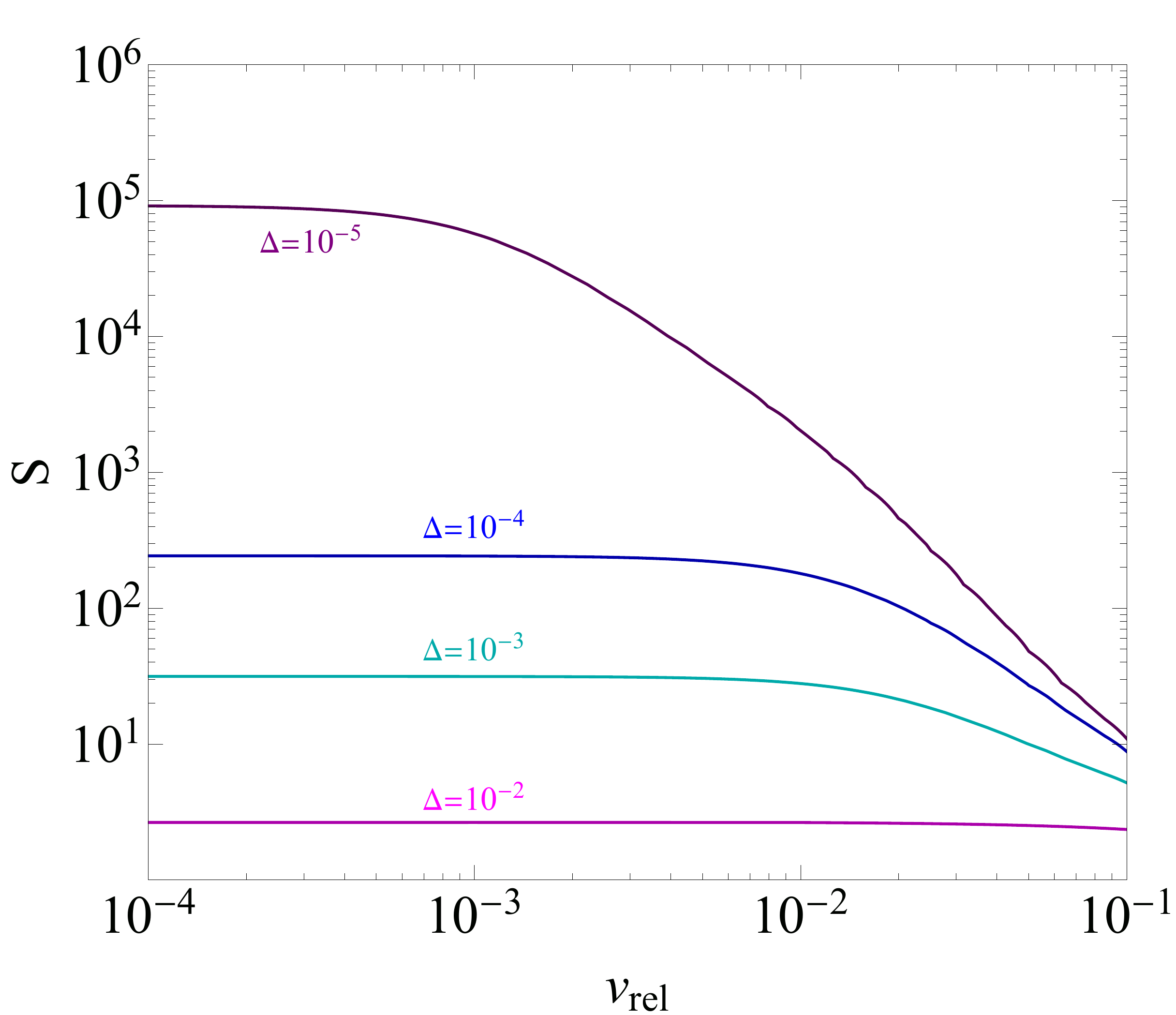} \,\,\,
\includegraphics[width=0.475\textwidth,clip]{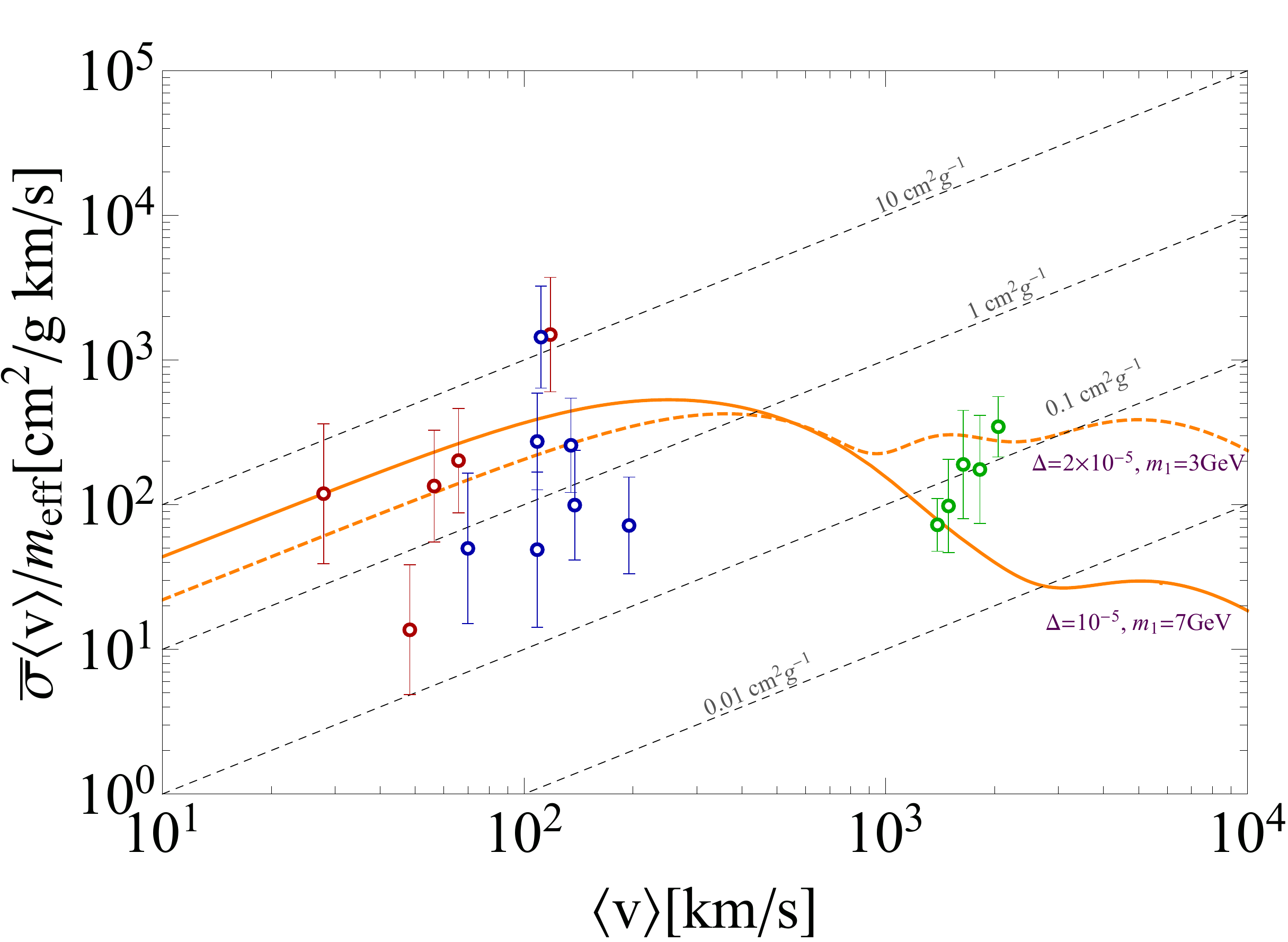} 
\end{center}
\caption[]{(Right) Sommerfeld factor as a function of dark matter relative velocity for different values of $\Delta\equiv 1-\beta/2$. (Left) Energy-transfer averaged co-scattering cross section per effective mass, $\overline{\sigma_V}/m_{\rm eff}$, in the presence of the $u$-channel resonance \cite{hmlee2}. Here, $m_{\rm eff}=2m_1(1+\beta)$. We took $(\Delta,m_1)=(10^{-5}, 7\,{\rm GeV}), (2\times 10^{-5}, 3\,{\rm GeV})$  in orange solid and dashed lines, respectively. For both plots, we set $\alpha=g^2/(4\pi)=0.1$. }
\label{fig:self}
\end{figure}

The co-scattering cross section for $\phi_1\phi_2\to \phi_1\phi_2$ is velocity-dependent due to the effectively light mediator in the $u$-channel. For instance, we consider the energy-transfer average of the co-scattering cross section by $\overline{\sigma_V}=\frac{\sqrt{\pi} v^3_0}{24}\langle\sigma_V v^3\rangle$ where $\sigma_V =\int d\Omega \sin^2\theta \frac{d\sigma}{d\Omega}$ is the viscosity cross section \cite{semi}, as shown in Fig.~\ref{fig:self}. 

We note that the $s$-channel contribution to $\phi_1\phi^*_1\to \phi_1\phi^*_1$ can be also enhanced at resonance for $m_2\simeq 2m_1$, similarly to the $u$-channel resonance. The $s$-channel total cross section is given by
\bea
\sigma_s=\frac{g^4m^2_1}{4\pi (m^2_1(4+v^2)-m^2_2)^2}\simeq \frac{g^4}{4\pi m^2_1 (v^2+8\Delta)^2}
\eea
where the width for $\phi_2$ is set to zero for $\Delta>0$, namely, $m_2<2m_1$. 
So, for $v\gtrsim \sqrt{8\Delta}$ and $\Delta\ll 1$, the cross section becomes $\sigma_s\simeq \frac{4\alpha}{9v^2}\, \frac{4\pi}{k^2}$. Therefore, for $v^2\sim \frac{4}{9}\alpha$, the $s$-channel resonance is comparable to the $u$-channel resonance at $\delta_0\simeq \frac{\pi}{2}$ in eq.~(\ref{selfcross}).
 
The mechanism for the $u$-channel resonance has been generalized to combinations of dark matter with different spins and parities and the case with more than two components of dark matter \cite{hmlee2}.

\section{Relic density for SRDM and dark photon portals}

\begin{figure}
\begin{center}
\includegraphics[width=0.8\textwidth,clip]{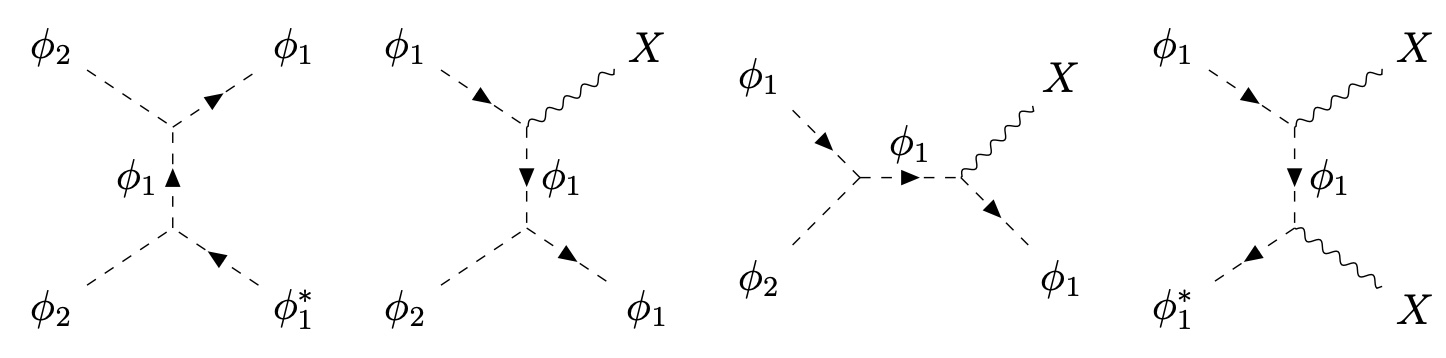} 
\end{center}
\caption[]{Feynman diagrams for dark matter annihilation in dark photon portals \cite{hmlee2}.}
\label{fig:ann}
\end{figure}

For self-resonant dark matter, as in the far left diagram in Fig.~\ref{fig:ann}, a heavier pair of $\phi_2$'s always annihilate into a lighter pair of $\phi_1$ and $\phi^*_1$, but dark matter abundances would be $Y_1= 2Y_2$, being unsuppressed \cite{hmlee2}. Thus, we need a depletion mechanism to reduce the abundance of the lighter dark matter particles or both.

We can make a simple extension of the self-resonant dark matter model in eq.~(\ref{SRDM}) with a dark photon $X_\mu$ where $X_\mu$ couples to the lighter dark matter $\phi_1$ from ${\cal L}=|D_\mu \phi_1|^2$ with $D_\mu\phi_1=(\partial_\mu-ig_X X_\mu)\phi_1$. Then, if $X_\mu$ decays into lighter particles such as the SM leptons, the heavier dark matter $\phi_2$ would decay by $\phi_2\to XX$ due to the one-loop tadpole diagram with $\phi_1$ in the loop \footnote{We thank Ryosuke Sato for pointing this out. }.
If $X_\mu$ is stable, we can forbid the loop-induced decay, $\phi_2\to XX$, kinematically, for $m_X>m_2/2$.  But, in this case, we also need to determine the relic abundance for $X$ carefully \cite{cline}.
Therefore, we need a mechanism for the stability of the heavier dark matter $\phi_2$. 

A simple solution for the dark matter stability is to add a new complex scalar $\phi'_1$ with equal mass as $\phi_1$ and introduce a $Z_2$ symmetry, as follows,
\bea
\phi_2\to -\phi_2,\qquad \phi_1\leftrightarrow \phi'_1.
\eea
Then, we extend the Lagrangian for self-resonant dark matter in eq.~(\ref{SRDM}) including the mass terms  and dark photon couplings \cite{work} to
\bea
{\cal L}&=&\frac{1}{2}(\partial_\mu\phi_2)^2+ |D_\mu\phi_1|^2+|D_\mu\phi'_1|^2-\frac{1}{2}m^2_2 \phi^2_2-m^2_1(|\phi_1|^2+|\phi'_1|^2) \nonumber \\
&&-2 gm_1 \phi_2 (|\phi_1|^2-|\phi'_1|^2)+\cdots
\eea
where $D_\mu\phi_{1,2}=(\partial_\mu-ig_X X_\mu)\phi_{1,2}$ and the ellipses are the quartic couplings for $\phi_1, \phi'_1$ and $\phi_2$, respecting the $Z_2$ symmetry, which are not important for depleting the lighter dark matter component. In this case, the heavier dark matter $\phi_2$ is stable because the one-loop tadpole diagrams are cancelled between $\phi_1$ and $\phi'_1$ thanks to their linear couplings to $\phi_2$ with opposite signs and the cancellation persists at all orders in perturbation theory. 
We also maintain the $u$-channel enhancement mechanism for the co-scattering processes, $\phi_1\phi_2\to \phi_1\phi_2$ and $\phi'_1\phi_2\to \phi'_1\phi_2$,  with $m_2\simeq 2m_1$, as discussed in the previous sections.

In the presence of dark photon couplings to the lighter dark matter components, $\phi_1$ and $\phi'_1$, we can deplete them sufficiently by the freeze-out mechanism.  For $g\gg g_X$, $\phi_2\phi_2\to \phi_1 \phi^*_1$ and a similar process for $\phi'_1$ annihilate until late times, so we can have a relation for the relic abundances by $Y_1\simeq Y'_1\simeq 2Y_2$ \cite{hmlee2}. In the opposite case with $g\ll g_X$, new annihilation channels, $\phi_1\phi^*_1\to XX$, as shown in the far right diagram in Fig.~\ref{fig:ann}, and a similar process for $\phi'_1$, are active until late times, so we get a hierarchy of the relic abundances by $Y_1\sim Y'_1\ll Y_2$ \cite{hmlee2}. 

The co-annihilation channel, $\phi_1\phi_2\to \phi_1 X$, as shown in the second and third diagrams of Fig.~\ref{fig:ann}, and a similar channel for $\phi'_1$, are subdominant during freeze-out for either limits, $g\gg g_X$ or  $g\ll g_X$, but they can be enhanced by the Sommerfeld effects when the dark matter velocity becomes small at later times in cosmology or at galaxies. Thus, if $\phi_1$ and $\phi'_1$ couple to the SM fermions through the dark photon portal, there are interesting signals for indirect and direct detections, due to the enhanced annihilation cross section at CMB and galaxies, as well as the boosted dark matter components, $\phi_1$ and $\phi'_1$, produced from the co-annihilations \cite{hmlee2}. 
Therefore, a more dedicated study on the constraints on the self-resonant dark matter model is demanding.

\section*{Acknowledgments}
The author thanks Seong-Sik Kim and Bin Zhu for collaboration during the projects and the organizers of the 19th Rencontres du Vietnam Theory Meeting Experiment (TMEX) 2023 for invitation and discussion during the workshop. 
The work is supported in part by Basic Science Research Program through the National Research Foundation of Korea (NRF) funded by the Ministry of Education, Science and Technology (NRF-2022R1A2C2003567 and NRF-2021R1A4A2001897).


\end{document}